\RequirePackage{fix-cm}
\documentclass[10pt,a4paper,twocolumn]{article}
\usepackage[a4paper,top=0.70in,bottom=0.65in,left=0.705in,right=0.705in]{geometry}
\usepackage[T1]{fontenc}
\usepackage[utf8]{inputenc}
\usepackage{times}
\usepackage{graphicx}
\usepackage{amsmath,amssymb,amsthm}
\usepackage{booktabs,tabularx,array}
\usepackage{tikz}
\usetikzlibrary{positioning,arrows.meta}
\usepackage{microtype}
\usepackage{placeins}
\usepackage{balance}
\AddToHook{shipout/after}[ccrankmixer/last-page-balance]{%
  \ifnum\PreviousTotalPages>1\relax
    \ifnum\ReadonlyShipoutCounter=\numexpr\PreviousTotalPages-1\relax
      \balance
      \typeout{CC-RankMixer: balancing final page \PreviousTotalPages.}%
    \fi
  \fi
}
\usepackage[font=small,labelfont=bf,labelsep=colon,justification=raggedright,singlelinecheck=false]{caption}

\usepackage{float}
\usepackage{needspace}
\usepackage{url}
\usepackage{xcolor}
\usepackage{hyperref}
\usepackage{titlesec}
\hypersetup{hidelinks,pdftitle={CC-RankMixer: View-Centered Training of Per-Token FFNs for Industrial Ranking},pdfauthor={Chao Ma, Shun Jia, Chuan Qin, Qiangsheng Bai, Heng Li, Yang Zhang, Yang Zhao, Jiawei Kang, Li Zhang, Qiang Fu}}
\newtheorem{proposition}{Proposition}
\titleformat{\section}{\fontsize{10.5}{12.5}\selectfont\bfseries}{\thesection}{0.55em}{}
\titleformat{\subsection}{\normalsize\bfseries}{\thesubsection}{0.55em}{}
\titleformat{\subsubsection}{\normalsize\itshape}{\thesubsubsection}{0.55em}{}
\titlespacing*{\section}{0pt}{10pt plus 2pt minus 1pt}{5pt}
\titlespacing*{\subsection}{0pt}{9pt plus 1pt minus 1pt}{4pt}
\begin{document}
\twocolumn[{%
\begin{@twocolumnfalse}
\noindent\includegraphics[width=79.4pt]{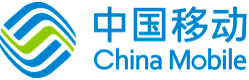}%
\hspace{15.4pt}\raisebox{5.1pt}{\includegraphics[width=107.5pt]{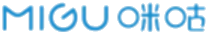}}\par
\vspace{16pt}
{\centering
{\fontsize{14.1}{16.5}\selectfont\bfseries CC-RankMixer: View-Centered Training of\\[2pt] Per-Token FFNs for Industrial Ranking\par}
\vspace{7pt}
{\fontsize{9}{11}\selectfont
Chao Ma$^{1,*}$, Shun Jia$^{1,2}$, Chuan Qin$^{3}$, Qiangsheng Bai$^{1}$, Heng Li$^{1}$\\[2pt]
Yang Zhang$^{1}$, Yang Zhao$^{1}$, Jiawei Kang$^{1}$, Li Zhang$^{1}$, Qiang Fu$^{1}$\par}
\vspace{3pt}
{\fontsize{8.6}{11}\selectfont
$^{1}$China Mobile / Migu Music Co., Ltd., Chengdu, China\\
$^{2}$Nankai University\\
$^{3}$Computer Network Information Center, Chinese Academy of Sciences\\
$^{*}$Corresponding author\par}
\vspace{18pt}
{\normalsize\bfseries Abstract\par}
\vspace{8pt}
\begin{minipage}{0.92\textwidth}
\fontsize{9.2}{11.05}\selectfont
Industrial ranking models need expressive feature interactions without increasing serving cost. RankMixer uses parameter-free token mixing and independent per-token feed-forward networks (P-FFNs). We propose CC-RankMixer, a view-centered training parameterization that couples P-FFNs through a shared weight center while preserving view-specific transformations through multiplicative offsets. The learned effective weights fold exactly into ordinary independent P-FFNs before deployment, preserving the deployed P-FFN graph and its theoretical inference computation. We derive cross-view gradient coupling and effective-weight updates to characterize how shared and specialized learning signals interact. Using the original ranking loss with Adagrad, CC-RankMixer improves both AUC and LogLoss over RankMixer on KuaiRand-27K, MovieLens-1M, and industrial music. On industrial music, AUC increases from 0.7119 to 0.7245 and LogLoss decreases from 0.6190 to 0.6103. A five-day online A/B test observes gains of 3.83\% in duration per play and 7.41\% in duration per user. Measured view CKA and cross-view gradient-cosine statistics show greater representation similarity and alignment, providing empirical support for coordinated learning across views.
\end{minipage}\par
\vspace{11pt}
}\noindent{\fontsize{9}{11}\selectfont \textbf{CCS Concepts:} Information systems $\rightarrow$ Recommender systems.\par
}\par
\vspace{9pt}
\end{@twocolumnfalse}
}]
\section{Introduction}
Industrial ranking systems must learn interactions among heterogeneous features under strict inference-latency and throughput constraints. RankMixer addresses this problem with parameter-free token rearrangement and independent per-token feed-forward networks (P-FFNs), concentrating interaction capacity in hardware-friendly dense computation~\cite{ref25}. We study a complementary question: \emph{how should the parameters of those independent FFNs be learned?}

Fully independent P-FFNs support specialization but provide no explicit coordination across tokens. A fully shared FFN couples all tokens to one transformation and can reduce accuracy~\cite{ref25}. Each mixed token combines one subspace from every tokenized input group. The resulting tokens are simultaneous latent views of the same example, potentially sharing predictive structure while retaining different information. This interpretation motivates partial parameter sharing; it does not imply independent modalities or preservation of every raw field.

We propose \textbf{CC-RankMixer} (Common--Complementary RankMixer), which expresses each P-FFN matrix through a shared trainable center and zero-initialized, view-specific multiplicative offsets. Biases use additive offsets. The effective parameters fold offline into ordinary independent P-FFNs, preserving their matrix shapes and theoretical inference computation. Training retains the original ranking objective without auxiliary view supervision or offset regularization.

We derive cross-view gradient coupling and the induced effective-weight update. A local SGD analysis separates an aggregate gradient contribution from a view-local, center-modulated contribution and distinguishes common-view from differential update directions. The reported experiments use Adagrad; the analysis characterizes the parameterization's local geometry rather than that optimizer's trajectory or a guaranteed accuracy benefit.

We compare four architectures on KuaiRand-27K, Taobao-UB, MovieLens-1M, and industrial music. CC-RankMixer improves both AUC and LogLoss over RankMixer on three datasets, with regressions on Taobao-UB. Configuration records explore token count and width, while training-loss summaries and validation curves show that better binary metrics need not coincide with lower recorded loss. Measured CKA and reported gradient statistics are consistent with greater cross-view representation similarity and gradient alignment; they do not establish causality. A formal five-day A/B test in the \emph{You May Like} scenario provides duration-focused online evidence.

Our contributions are:
\begin{itemize}
\item A view-centered training parameterization that couples simultaneous token views while retaining separate effective P-FFNs.
\item Cross-view gradient and effective-update analysis, together with exact offline folding and backbone cost accounting.
\item Offline ranking, configuration, training-behavior, and online results, with measured representation and gradient diagnostics.
\end{itemize}

\section{Related Work}
\noindent\textbf{Feature interaction.} Prior architectures include factorization machines~\cite{ref_fm}, Wide\&Deep~\cite{ref3}, DeepFM~\cite{ref6}, DCN~\cite{ref_dcn}, xDeepFM~\cite{ref_xdeepfm}, FiBiNET~\cite{ref_fibinet}, AutoInt~\cite{ref16}, and Wukong~\cite{ref23}. DIN and DIEN emphasize adaptive and evolving user-interest representations rather than our parameterization of independent per-token FFNs~\cite{ref_din,ref_dien}. MLP-Mixer explores token/channel mixing in vision~\cite{ref_mlp_mixer}, whereas RankMixer~\cite{ref25} uses parameter-free token rearrangement and per-token FFNs. We retain its mixing operator and deployed P-FFN graph while changing the organization of training parameters.

\noindent\textbf{Sharing and specialization.} Shared--private representations~\cite{ref2}, cross-stitch units~\cite{ref_crossstitch}, MMoE~\cite{ref10}, and PLE~\cite{ref17} combine common and specialized components across tasks or domains. STAR~\cite{ref15} is closest in parameter form: it modulates a shared body with domain-specific multiplicative factors. CC-RankMixer applies related composition to simultaneous subspace views of one input, without a domain indicator. Multiplicative sharing itself is established prior work; our contribution concerns view construction, coupled optimization, folding to independent P-FFNs, and comparison with RankMixer.

\noindent\textbf{Adapters and mixtures.} LoRA~\cite{ref8} learns low-rank updates for a frozen base, whereas our full-shaped offsets and centers are trained jointly from scratch. An activation-space shared--private model executes two branches, $F_s(m_h)+F_h(m_h)$~\cite{ref2}; sparse MoE models combine nonlinear expert outputs~\cite{ref5,ref14}. CC-RankMixer instead executes one effective FFN per view. These methods are conceptual comparisons; RankMixer is the central empirical baseline, with DNN and Wukong providing architecture-level reference points.

\section{Method}
\label{sec:method}
\subsection{Overview and RankMixer Backbone}
\begin{figure*}[!t]
\centering
\includegraphics[width=\textwidth]{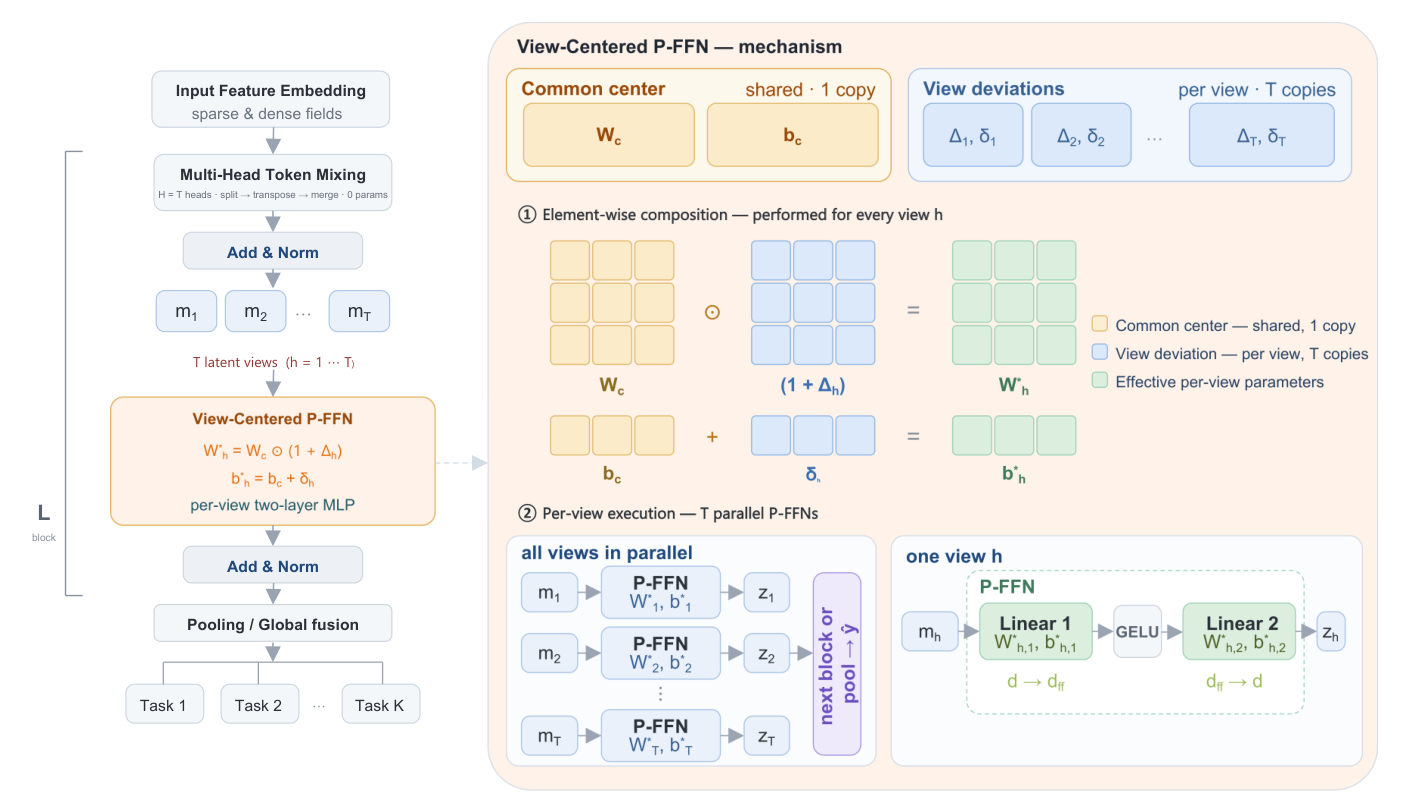}
\caption{CC-RankMixer architecture and the view-centered P-FFN mechanism. The shared center and view-specific offsets are composed during training and folded to per-view P-FFNs for inference.}
\label{fig:overview}
\end{figure*}

CC-RankMixer retains RankMixer's mixing operator and residual backbone (Figure~\ref{fig:overview}) and changes the training parameterization of its P-FFNs:
\begin{equation}\label{eq:pipeline}
\bigl[X\xrightarrow{\mathrm{Mix}}M
\xrightarrow{\mathrm{CC\mbox{-}PFFN}}Z\bigr]^{\times L}
\longrightarrow\hat y.
\end{equation}
Mixed tokens act as latent views; their effective FFN weights are trained through a common center and view-specific offsets, then folded before inference.

Feature tokenization produces $T$ tokens of width $D$~\cite{ref25}. With $H=T$ mixing heads, each token is divided into $T$ sub-blocks:
\begin{align}
X&=[x_1,\ldots,x_T],\qquad x_i\in\mathbb R^D,\label{eq:tokens}\\
x_i&=[x_i^{(1)},\ldots,x_i^{(T)}],\qquad
x_i^{(h)}\in\mathbb R^{D/T}.\label{eq:split}
\end{align}
The split--transpose--merge operator reassembles them as
\begin{equation}\label{eq:mixed}
m_h=\operatorname{Concat}(x_1^{(h)},\ldots,x_T^{(h)}),
\qquad M=[m_1,\ldots,m_T].
\end{equation}
This operation uses only reshapes and a transpose. Each mixed token passes through its own two-layer P-FFN,
\begin{equation}\label{eq:pffn}
u_h=\phi(m_hW_{1,h}+b_{1,h}),\qquad
z_h=u_hW_{2,h}+b_{2,h},
\end{equation}
within a block using residual connections and layer normalization~\cite{ref1,ref7}. Block outputs are pooled and fed to task heads.

\subsection{Latent Views and Parameter Sharing}
\label{sec:latent}
Each $m_h$ receives a sub-block from every \emph{tokenized} input, but only $D/T$ coordinates from each. We therefore interpret the mixed tokens as simultaneous partial views of one example. This does not guarantee that tokenization preserves all raw-field information.

Two hypotheses motivate sharing. \emph{Shared regularities:} a common example and target may yield correlated useful mappings. \emph{View-specific information:} distinct subspaces may provide nonredundant information. Complementarity requires empirical combination probes; disjoint coordinates alone do not establish it. The limiting designs are
\begin{align}
z_h&=F_{\mathrm{shared}}(m_h) &&\text{(fully shared)},\label{eq:shared}\\
z_h&=F_h(m_h) &&\text{(independent P-FFNs)}.\label{eq:indep}
\end{align}
RankMixer's ablations report reduced predictive quality under full FFN sharing~\cite{ref25}. Independent P-FFNs permit specialization but do not explicitly couple their optimization. We couple updates through shared parameters while retaining separate effective FFNs. Section~\ref{sec:mechanism} examines measured representation similarity and gradient alignment; these diagnostics do not by themselves prove predictive complementarity.

\subsection{View-Centered P-FFN}
\label{sec:center}
A common center $(W_c,b_c)$ and view-specific offsets define
\begin{equation}\label{eq:param}
W_h^\star=W_c\odot(\mathbf1+\Delta_h),\qquad
b_h^\star=b_c+\delta_h,
\end{equation}
where $\odot$ denotes element-wise multiplication. Zero initialization, $\Delta_h=0$ and $\delta_h=0$, gives all views the same initial transformation. Both layers use
\begin{equation}
\begin{aligned}
W_{k,h}^\star &=W_{k,c}\odot(\mathbf1+\Delta_{k,h}),\\
b_{k,h}^\star &=b_{k,c}+\delta_{k,h},\quad k\in\{1,2\}.
\end{aligned}
\end{equation}
Each view executes one FFN, $z_h=\operatorname{PFFN}(m_h;W_h^\star,b_h^\star)$.

For one weight matrix, let $G_h=\partial\mathcal L/\partial W_h^\star$. The chain rule gives
\begin{equation}\label{eq:grad}
\nabla_{W_c}\mathcal L=\sum_jG_j\odot(\mathbf1+\Delta_j),
\qquad
\nabla_{\Delta_h}\mathcal L=G_h\odot W_c.
\end{equation}
Likewise, $\nabla_{b_c}\mathcal L=\sum_j\nabla_{b_j^\star}\mathcal L$ and $\nabla_{\delta_h}\mathcal L=\nabla_{b_h^\star}\mathcal L$. The center aggregates view contributions; each offset receives its own view's center-modulated gradient. The offset gradient is not automatically a residual after subtracting a common direction. Variance-reduction claims would require additional assumptions on alignment and noise covariance.

\subsection{Effective Weight Update Analysis}
\label{sec:update}
The gradient identities hold for any optimizer. To isolate the parameterization's effect, we analyze plain SGD with center and offset learning rates $\eta_c,\eta_\Delta$. The experiments use Adagrad~\cite{ref_adagrad}; its adaptive trajectory is outside this analysis. Define
\begin{equation}\label{eq:gradagg}
A=\sum_{j=1}^{T}G_j\odot(\mathbf1+\Delta_j).
\end{equation}
An SGD step gives $W_c'=W_c-\eta_cA$ and $\Delta_h'=\Delta_h-\eta_\Delta(G_h\odot W_c)$. Expanding the effective weight yields the exact finite-step identity
\begin{equation}\label{eq:exactupdate}
\begin{aligned}
W_h^{\star\prime}-W_h^\star
={}&-\eta_cA\odot(\mathbf1+\Delta_h)
-\eta_\Delta W_c^{\odot2}\odot G_h\\
&+\eta_c\eta_\Delta A\odot W_c\odot G_h.
\end{aligned}
\end{equation}
At zero-offset initialization, omitting the order-$\eta_c\eta_\Delta$ term gives
\begin{equation}\label{eq:effectiveupdate}
\boxed{\Delta W_h^\star\approx
-\eta_c\sum_{j=1}^{T}G_j
-\eta_\Delta W_c^{\odot2}\odot G_h.}
\end{equation}
The first term couples every view to the aggregate gradient; the second preserves a view-local update scaled by the squared center weights. Independent-P-FFN SGD instead gives $\Delta W_h=-\eta G_h$.

For one scalar coordinate, collect the view gradients into $\mathbf g=(g_1,\ldots,g_T)^\top$. At initialization, $\Delta\mathbf w\approx-P\mathbf g$, with
\begin{equation}\label{eq:preconditioner}
P=\eta_c\mathbf1\mathbf1^\top+\eta_\Delta w_c^2I_T.
\end{equation}
The common-view direction has eigenvalue $\eta_cT+\eta_\Delta w_c^2$; orthogonal difference directions have eigenvalue $\eta_\Delta w_c^2$. Their distinct update scales motivate a center-learning-rate normalization control as $T$ changes. This local geometry establishes neither improved convergence nor generalization, and its scalar learning rates do not represent Adagrad's accumulated coordinate-wise adaptation.

\subsection{Training Objective}
\label{sec:objectives}
Training uses the backbone's original ranking objective:
\begin{equation}\label{eq:loss}
\boxed{\mathcal L_{\mathrm{CC}}=\mathcal L_{\mathrm{rank}}.}
\end{equation}
Its targets include click and duration modeling, without CC-specific auxiliary view supervision, center constraints, or offset penalties. Reported view-head parameters do not imply an additional loss. Head connectivity, shared initialization, and optimizer effects remain possible alternative explanations requiring controls. Training and validation losses retain their source definitions; neither replaces binary-task LogLoss. Duration-like targets on Taobao-UB and MovieLens-1M are proxies, not observed listening time.

\subsection{Inference-Time Folding and Cost}
\label{sec:fold}
\begin{figure}[!tbp]
\centering
\begin{tikzpicture}[>=Stealth,
 every node/.style={font=\fontsize{7}{8.4}\selectfont,align=center},
 b/.style={draw=gray!45,rounded corners=2pt,minimum height=0.57cm,inner sep=2pt}]
\node[b,fill=orange!8,minimum width=2.55cm] (center) at (-1.45,0) {Shared center\\$W_c,b_c$};
\node[b,fill=blue!7,minimum width=2.55cm] (offset) at (1.45,0) {View-specific offsets\\$\Delta_h,\delta_h$};
\node[b,fill=green!7,minimum width=5.5cm] (eff) at (0,-1.02) {$W_h^\star=W_c\odot(1+\Delta_h),\quad b_h^\star=b_c+\delta_h$};
\node[b,fill=gray!8,minimum width=5.5cm] (fold) at (0,-2.02) {Offline folding $\rightarrow$ ordinary independent P-FFN};
\draw[->,gray!70] (center.south) -- ([xshift=-12pt]eff.north);
\draw[->,gray!70] (offset.south) -- ([xshift=12pt]eff.north);
\draw[->,gray!70] (eff.south) -- (fold.north);
\end{tikzpicture}
\caption{View-centered weight composition and offline folding. The deployed inference path executes ordinary per-token FFNs.}
\label{fig:folding}
\end{figure}
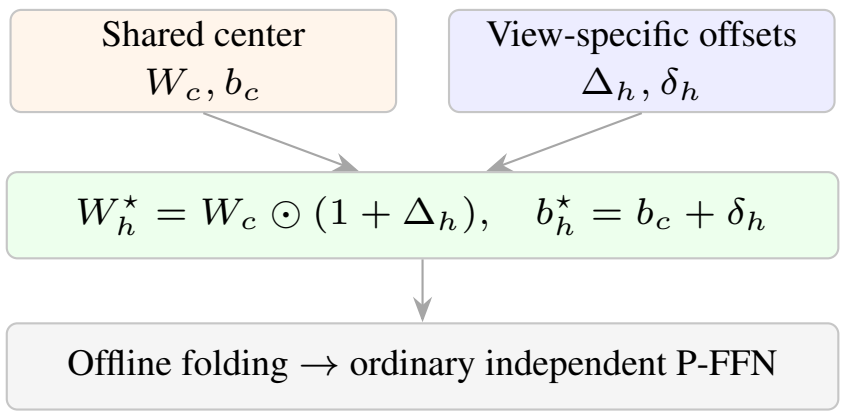

Before serving, we materialize the effective weights for each layer and view:
\begin{equation}\label{eq:fold}
W_h=W_c\odot(\mathbf1+\Delta_h),\qquad b_h=b_c+\delta_h.
\end{equation}
They are stored under ordinary per-token FFN parameter names, eliminating online weight composition (Figure~\ref{fig:folding}).

\begin{proposition}[Exactness of folding]\label{prop:folding}
For the two-layer P-FFN $F$ in Eq.~\eqref{eq:pffn}, every view $h$ and input $m_h$ satisfy
\begin{equation}
F(m_h;W_h^\star,b_h^\star)=F(m_h;W_h,b_h),
\end{equation}
using Eqs.~\eqref{eq:param} and~\eqref{eq:fold}. If the remaining scoring graph is unchanged, this equality extends to final predictions, up to floating-point effects.
\end{proposition}
\begin{proof}
Substitution gives identical first-layer pre-activations,
$m_hW_{1,h}+b_{1,h}=m_h[W_{1,c}\odot(\mathbf1+\Delta_{1,h})]+b_{1,c}+\delta_{1,h}$.
The same activation and second-layer substitution give identical FFN outputs. Equality propagates through unchanged backbone, pooling, and head operations. It is exact in real arithmetic; deployment precision and operation ordering govern numerical agreement.
\end{proof}

A P-FFN weight matrix of shape $D\times D'$ requires $TDD'$ parameters in RankMixer and $(T+1)DD'$ during CC training. Thus the relative increment is
\begin{equation}\label{eq:params}
\frac{(T+1)DD'-TDD'}{TDD'}=\frac1T,
\end{equation}
or 12.5\% at $T=8$ and 6.25\% at $T=16$. This applies to both FFN layers. Folding retains only $T$ effective weight sets. Embeddings and separate heads are outside this calculation; their parameters are not asserted to fold away.

\begin{table}[H]
\centering
\caption{Inference-path equivalence after offline P-FFN folding.}
\label{tab:complexity}
\begingroup
\footnotesize
\setlength{\tabcolsep}{3pt}
\renewcommand{\arraystretch}{1.16}
\begin{tabularx}{\columnwidth}{@{}>{\raggedright\arraybackslash}X
  >{\raggedright\arraybackslash}p{0.37\columnwidth}@{}}
\toprule
Backbone property & Folded CC-RankMixer \\
\midrule
Token mixing operator & Unchanged \\
Number of P-FFNs & $T$ (unchanged) \\
P-FFN shapes / FLOPs & Unchanged \\
P-FFN activations & Unchanged \\
Grouped GEMM & Preserved \\
Online weight composition & Not required \\
Deployed P-FFN parameter count & Unchanged \\
\bottomrule
\end{tabularx}
\endgroup
\vspace{2pt}
{\scriptsize\raggedright Relative to RankMixer; embedding and task-head parameters are excluded.\par}
\end{table}

Training adds weight-composition operations and stores centers, offsets, optimizer state, and temporary tensors. Their runtime and memory cost depend on implementation, not parameter counts alone. Table~\ref{tab:complexity} describes theoretical backbone costs; Table~\ref{tab:implementation} reports the separate full-model training inventory.

\section{Experiments}
\label{sec:experiments}
We examine ranking effectiveness, configuration sensitivity, training behavior, and online utility. We also evaluate measured view similarity and gradient alignment.

\subsection{Experimental Setup}
\label{sec:setup}
\textbf{Datasets and targets.} Table~\ref{tab:datasets} summarizes three public datasets---KuaiRand-27K~\cite{ref_kuairand}, Taobao-UB~\cite{ref_taobao}, and MovieLens-1M~\cite{ref_movielens}---together with an industrial music dataset. Their binary targets are click, purchase, rating $>3$, and click, respectively. Observed play duration is available in the video and music records; the other datasets are evaluated primarily on their binary targets. Duration-like proxy labels also appear in the source experiment records for Taobao-UB and MovieLens-1M; their construction is not documented, and they are not treated as measured play duration. Table~\ref{tab:datasets} records the reported time-based evaluation protocols.

\begin{table*}[!t]
\centering
\caption{Public and industrial datasets: scale, targets, and evaluation protocols.}
\label{tab:datasets}
\begingroup
\small
\renewcommand{\arraystretch}{1.16}
\setlength{\tabcolsep}{3pt}
{\bfseries Panel A: Dataset scale}\par\vspace{4pt}
\begin{tabular*}{\textwidth}{@{\extracolsep{\fill}}llrrrr@{}}
\toprule
Dataset & Domain & Users & Items & Samples & Fields \\
\midrule
KuaiRand-27K & Video & 27,285 & 32,038,725 & 322,278,385 & 40 \\
Taobao-UB & Commerce & 987,994 & 4,162,024 & 100,150,807 & 8 \\
MovieLens-1M & Movies & 6,040 & 3,706 & 1,000,209 & 11 \\
Industrial music & Music & 787,526 & 608,400 & 31,017,185 & 58 \\
\bottomrule
\end{tabular*}
\par\vspace{9pt}
{\bfseries Panel B: Targets and chronological evaluation}\par\vspace{4pt}
\begin{tabularx}{\textwidth}{@{}
>{\raggedright\arraybackslash}p{2.05cm}
>{\raggedright\arraybackslash}p{2.20cm}
>{\raggedright\arraybackslash}p{1.75cm}
>{\raggedright\arraybackslash}X@{}}
\toprule
Dataset & Binary target & Duration signal & Train / evaluation protocol \\
\midrule
KuaiRand-27K & Click (\texttt{is\_click}) & Play duration &
Train: date $<$ 2022-05-04;\newline evaluation: date $\geq$ 2022-05-04. \\
\addlinespace[3pt]
Taobao-UB & Purchase (\texttt{is\_buy}) & Duration proxy &
Train: 2017-11-25--2017-12-02;\newline evaluation: 2017-12-03 (last day). \\
\addlinespace[3pt]
MovieLens-1M & Rating $>3$ (\texttt{is\_pos}) & Duration proxy &
Chronological cutoff; final 20\% for evaluation. \\
\addlinespace[3pt]
Industrial music & Click & Play duration &
Train: 2026-11-01--2026-11-10 (10 days);\newline evaluation: 2026-11-11 (1 day). \\
\bottomrule
\end{tabularx}
\endgroup
\vspace{3pt}
{\footnotesize\raggedright Reported dataset counts and field dimensions. Constructed duration-proxy labels in Taobao-UB and MovieLens-1M are distinct from observed play duration; their precise construction is not documented. Chronological cutoffs are shown as recorded.\par}
\end{table*}

\textbf{Models and training.} We compare DNN, Wukong~\cite{ref23}, RankMixer~\cite{ref25}, and CC-RankMixer. The two mixers use Adagrad~\cite{ref_adagrad}, learning rate 0.005, training batch 1024, and evaluation batch 4096. Their default backbone is $T=16$, $D=768$, $L=2$ mixer blocks. Explicitly labeled sweeps override this default. An additional parameter inventory is given in Table~\ref{tab:implementation}.

Known or estimated non-embedding components are of order $10^8$: DNN's specified hidden matrices contain 151.7M weights; Wukong's principal blocks and head are estimated at 102M under a flattened-MLP layout; the mixer remainders are approximately 150M/160M. These counting scopes do not establish exact parameter matching. CC adds about 10M training parameters in view-centered FFNs and view heads (0.85\% of RankMixer's total). These counts do not determine head connectivity or serving treatment.

\textbf{Metrics.} AUC and LogLoss measure each binary task; the industrial sweep also reports MSE. Training and validation losses retain their source labels and are not substituted for binary LogLoss. Results are individual recorded runs, without repeated-seed uncertainty. Identical backbone configurations do not establish matched preprocessing, training budgets, or checkpoint selection.

\subsection{Overall Ranking Performance}
\label{sec:overall}
Table~\ref{tab:overall} reports all four models. CC-RankMixer improves \emph{both AUC and LogLoss} over RankMixer on KuaiRand-27K, MovieLens-1M, and industrial music. AUC increases by 0.0079, 0.0024, and 0.0126, while LogLoss decreases by 0.0090, 0.0022, and 0.0087. Taobao-UB regresses on both metrics, so the benefit is dataset-dependent.

\begin{table*}[!t]
\centering
\caption{Ranking performance across three public benchmarks and industrial music.}
\label{tab:overall}
\begingroup
\small
\renewcommand{\arraystretch}{1.18}
\setlength{\tabcolsep}{3pt}
\newcommand{\ResultPanel}[1]{%
\begin{tabularx}{\textwidth}{@{}>{\raggedright\arraybackslash}p{2.65cm}*{4}{>{\centering\arraybackslash}X}@{}}
\toprule
Model & \shortstack{KuaiRand-\\27K} & Taobao-UB & \shortstack{MovieLens-\\1M} & \shortstack{Industrial\\music} \\
\midrule
#1
\bottomrule
\end{tabularx}%
}
\par\vspace{3pt}
{\bfseries Panel A: AUC $\uparrow$}\par\vspace{4pt}
\ResultPanel{
DNN & \textbf{0.8145} & 0.8156 & 0.7784 & 0.7160 \\
Wukong & 0.7965 & \textbf{0.8664} & \textbf{0.7880} & 0.7119 \\
\midrule
RankMixer & 0.8052 & 0.8595 & 0.7856 & 0.7119 \\
\textbf{CC-RankMixer} & 0.8131 & 0.8557 & \textbf{0.7880} & \textbf{0.7245} \\
\midrule
CC $-$ RankMixer & +0.0079 & -0.0038 & +0.0024 & +0.0126 \\
}
\par\vspace{9pt}
{\bfseries Panel B: LogLoss $\downarrow$}\par\vspace{4pt}
\ResultPanel{
DNN & \textbf{0.4970} & 0.2157 & 0.5596 & 0.6215 \\
Wukong & 0.5219 & 0.1646 & 0.5730 & 0.6211 \\
\midrule
RankMixer & 0.5079 & \textbf{0.1645} & 0.5539 & 0.6190 \\
\textbf{CC-RankMixer} & 0.4989 & 0.1824 & \textbf{0.5517} & \textbf{0.6103} \\
\midrule
CC $-$ RankMixer & -0.0090 & +0.0179 & -0.0022 & -0.0087 \\
}
\endgroup
\vspace{3pt}
{\footnotesize\raggedright Boldface marks best scores (including ties). Differences denote CC-RankMixer minus RankMixer; positive AUC and negative LogLoss favor CC. Both mixers use $T=16$, $D=768$, $L=2$. Values are individual runs. Training-budget and checkpoint alignment remain unverified; Table~\ref{tab:scale} is a separate record set.\par}
\end{table*}

CC leads both industrial metrics and MovieLens-1M LogLoss, tying Wukong's MovieLens-1M AUC at the reported precision. DNN leads KuaiRand-27K; Wukong and RankMixer lead Taobao-UB AUC and LogLoss, respectively. The validation-AUC summaries agree with the mixer entries in Table~\ref{tab:overall}.

\subsection{Sensitivity to Token Number and Width}
\label{sec:scale}
Table~\ref{tab:scale} varies $T\in\{4,8,16\}$ at $D=256$, and $D\in\{256,384,768\}$ at $T=16$. The shared configuration $(16,256)$ appears once. At fixed $D=256$, recorded AUC rises from 0.700777 to 0.744190 and 0.747555 as $T$ increases, while LogLoss and MSE decrease. At $T=16$, wider configurations also have higher AUC and lower errors; $(16,768)$ is best among these five records.

\begin{table*}[!t]
\centering
\caption{Configuration sensitivity of CC-RankMixer on the industrial music dataset.}
\label{tab:scale}
\begingroup
\small
\renewcommand{\arraystretch}{1.18}
\setlength{\tabcolsep}{4pt}
\begin{tabularx}{\textwidth}{@{}*{6}{>{\centering\arraybackslash}X}@{}}
\toprule
$T$ & $D$ & AUC $\uparrow$ & LogLoss $\downarrow$ & MSE $\downarrow$ & Steps \\
\midrule
4 & 256 & 0.700777 & 0.656027 & 17213.72 & 363684 \\
8 & 256 & 0.744190 & 0.622070 & 15884.26 & 363684 \\
16 & 256 & 0.747555 & 0.600594 & 15868.14 & 363684 \\
16 & 384 & 0.754892 & 0.585494 & 15686.82 & 384804 \\
16 & 768 & 0.776716 & 0.498900 & 15541.73 & 366602 \\
\bottomrule
\end{tabularx}
\endgroup
\vspace{3pt}
{\footnotesize\raggedright All rows are separately recorded CC-RankMixer runs. Steps are the reported Global Steps. Training budgets vary, and these runs are not aligned with the main comparison in Table~\ref{tab:overall}; the table describes observed configuration performance.\par}
\end{table*}

Training steps vary, and the sweep is not aligned with the main comparison in data windows or preprocessing. In particular, the CC AUC values 0.7245 in Table~\ref{tab:overall} and 0.776716 in Table~\ref{tab:scale} are separate records, not a cross-record gain. This descriptive sweep establishes neither a controlled scaling law nor a scale-dependent advantage over RankMixer.

\subsection{Training and Validation Behavior}
\label{sec:training}
Table~\ref{tab:training} reports training-loss summaries. CC has lower loss on Taobao-UB and higher loss on the other datasets, the opposite of their binary-metric ordering. Loss magnitudes are interpretable only within each dataset and do not establish convergence speed.

\begin{table}[!htbp]
\centering
\caption{Reported training-loss summaries at $(T,D)=(16,768)$.}
\label{tab:training}
\begingroup\small
\renewcommand{\arraystretch}{1.15}
\begin{tabularx}{\columnwidth}{@{}>{\raggedright\arraybackslash}X
>{\centering\arraybackslash}p{0.235\columnwidth}
>{\centering\arraybackslash}p{0.235\columnwidth}@{}}
\toprule
Dataset & RankMixer & CC-RankMixer \\
\midrule
KuaiRand-27K & 0.7770 & 0.8406 \\
Taobao-UB & 1.1157 & 0.4866 \\
MovieLens-1M & 1.0666 & 1.1806 \\
Industrial music & 1.2281 & 1.2895 \\
\bottomrule
\end{tabularx}
\endgroup
\vspace{2pt}
{\scriptsize\raggedright These are the recorded training losses, not binary
LogLoss; checkpoint and objective normalization details were not provided.\par}
\end{table}

\begin{figure}[!htb]
\centering
\includegraphics[width=\columnwidth]{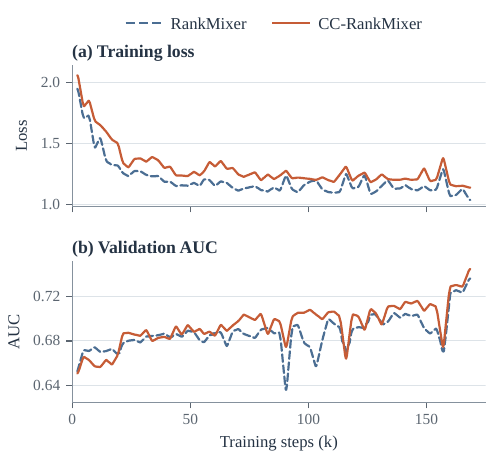}
\caption{Training-loss and validation-AUC trajectories. Curves are digitized from the supplied plot; their coordinates are approximate.}
\label{fig:convergence}
\end{figure}

The trajectories in Figure~\ref{fig:convergence} show an overall decline in training loss despite local fluctuations. CC-RankMixer generally has a higher displayed training loss, while its validation AUC is higher over much of the later trajectory. The digitized curves illustrate training behavior; their endpoints are not substituted for the independently reported summary values in Tables~\ref{tab:overall} and~\ref{tab:training}.

\begin{figure*}[!t]
\centering
\includegraphics[width=.88\textwidth]{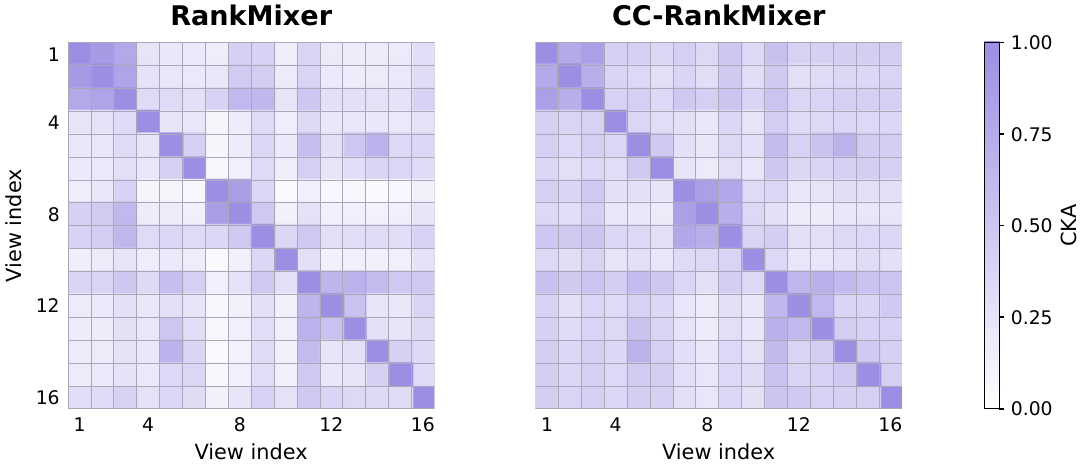}
\caption{Pairwise linear CKA for RankMixer and CC-RankMixer at $T=16$ ($N=5{,}000$), displayed using a common 0--1 color scale. Cell annotations are omitted for legibility; the corresponding summary statistics are in Table~\ref{tab:mechanism}.}
\label{fig:cka}
\end{figure*}
\subsection{Representation Similarity and Gradient Alignment}
\label{sec:mechanism}
We evaluate pairwise linear centered kernel alignment (CKA)~\cite{ref_cka} across $T=16$ latent-view representations using 5,000 examples. The mean off-diagonal CKA rises from 0.3017 for RankMixer to 0.3937 for CC-RankMixer (+30.49\% relative); the minimum pair similarity rises from 0.052 to 0.168. Figure~\ref{fig:cka} directly compares the \emph{two measured} $16\times16$ CKA matrices on a shared 0--1 color scale while their aggregate statistics are reported in Table~\ref{tab:mechanism}. Cluster structures remain visible in both models. Higher CKA indicates more common representational structure, but does not on its own establish predictive complementarity.

\begin{figure}[!htb]
\centering
\includegraphics[width=\columnwidth]{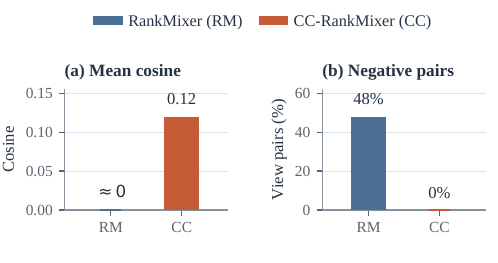}
\caption{Reported cross-view gradient-cosine statistics: mean cosine is approximately $0$ vs. $0.12$, and negative-cosine view-pair ratio is $48\%$ vs. $0\%$ (RankMixer vs. CC-RankMixer). The coordinate convention and uncertainty intervals were not reported.}
\label{fig:gradient}
\end{figure}

Figure~\ref{fig:gradient} visualizes the reported cross-view gradient-cosine statistics. Their precise coordinate convention was not supplied, so they should not be equated with the optimizer-specific effective-weight updates derived under SGD. The observed alignment is consistent with the proposed shared-center mechanism but does not establish a causal effect on ranking accuracy.

\begin{table}[H]
\centering
\caption{Cross-view CKA and gradient statistics ($T=16$).}
\label{tab:mechanism}
\begingroup
\small
\renewcommand{\arraystretch}{1.13}
\setlength{\tabcolsep}{3pt}
\begin{tabularx}{\columnwidth}{@{}>{\raggedright\arraybackslash}X
>{\centering\arraybackslash}p{0.235\columnwidth}
>{\centering\arraybackslash}p{0.235\columnwidth}@{}}
\toprule
Metric & RankMixer & CC-RankMixer \\
\midrule
\multicolumn{3}{@{}l}{\emph{Representation alignment}} \\
Mean off-diagonal CKA & 0.3017 & \textbf{0.3937} \\
Minimum pairwise CKA & 0.052 & 0.168 \\
Mean CKA fold std & 0.0079 & 0.0090 \\
Maximum CKA fold std & 0.0312 & 0.0344 \\
\addlinespace[2pt]
\multicolumn{3}{@{}l}{\emph{Gradient alignment}} \\
Mean gradient cosine & $\approx 0.00$ & 0.12 \\
Negative-cosine pair ratio & 48\% & 0\% \\
\bottomrule
\end{tabularx}
\endgroup
\vspace{2pt}
{\scriptsize\raggedright CKA uses 5,000 examples. Gradient confidence intervals and
coordinate conventions were not supplied; these metrics are not claims of
raw-task-gradient conflict removal.\par}
\end{table}

\Needspace{8\baselineskip}
\subsection{Industrial Online A/B Test}
\label{sec:online}
We conducted a formal A/B experiment in the \emph{You May Like} music recommendation scenario. Traffic was split by user cohort, with RankMixer serving the control cohort and CC-RankMixer the treatment cohort. Both arms used $(T,D)=(16,768)$. Over the five-day observation period, cumulative user-level results show increases of 3.83\% in duration per play and 7.41\% in duration per user, with two-sided $p$-values of 0.1169 and 0.1575 (Table~\ref{tab:online}).

\begin{table}[H]
\centering
\caption{Five-day online A/B results in \emph{You May Like}.}
\label{tab:online}
\begingroup
\fontsize{8.5}{10.5}\selectfont
\setlength{\tabcolsep}{1.2pt}
\renewcommand{\arraystretch}{1.18}
\begin{tabular*}{\columnwidth}{@{\extracolsep{\fill}}lrrrr@{}}
\toprule
Metric & RankMixer & CC-RankMixer & Lift & $p$ \\
\midrule
Duration / play (s) & 166.49 & 172.88 & +3.83\% & 0.1169 \\
Duration / user (s) & 1437.69 & 1544.17 & +7.41\% & 0.1575 \\
\bottomrule
\end{tabular*}
\endgroup
\vspace{1.5pt}
{\scriptsize\raggedright User-cohort traffic split, cumulative user-level statistics over five days. Operational threshold $p<0.20$ (neither outcome reaches $p<0.05$).\par}
\end{table}

\begin{table}[!htbp]
\centering
\caption{Training setup and approximate parameter inventory.}
\label{tab:implementation}
\begingroup
\fontsize{8.6}{10.0}\selectfont
\setlength{\tabcolsep}{2.7pt}
\renewcommand{\arraystretch}{1.07}
\begin{tabularx}{\columnwidth}{@{}>{\raggedright\arraybackslash}Xrr@{}}
\toprule
Setting & RankMixer & CC-RankMixer \\
\midrule
\multicolumn{3}{@{}l}{\itshape Training configuration} \\
Backbone $(T,D,L)$ & $(16,768,2)$ & $(16,768,2)$ \\
Optimizer & Adagrad & Adagrad \\
Learning rate & 0.005 & 0.005 \\
Train / eval batch & 1024 / 4096 & 1024 / 4096 \\
Objective & $\mathcal L_{\mathrm{rank}}$ & $\mathcal L_{\mathrm{rank}}$ \\
\addlinespace[1.8pt]
\multicolumn{3}{@{}l}{\itshape Parameters (approximate)} \\
Total & 1.18B & 1.19B \\
Embedding & 1.03B & 1.03B \\
Non-embedding & 150M & 160M \\
\bottomrule
\end{tabularx}
\endgroup
\vspace{3pt}
{\fontsize{7.7}{9}\selectfont\raggedright Counts describe training models, not folded serving models. Non-embedding entries are differences, not per-dataset measurements.\par}
\end{table}

\textbf{Operational criterion.} Experimental traffic was constrained and relatively few users participated. The business uses $p<0.20$ to balance the cost and delay of extending an experiment against greater decision uncertainty. A smaller sample does not strengthen the evidence. Both outcomes meet this criterion, while neither meets $p<0.05$. The experiment ended when the operational launch criterion was reached; five days is the realized observation period. These are the $p$-values reported at stopping; a sequential-testing correction is not documented.

Separately monitored guardrails require review before broad deployment claims. Serving latency under matched hardware is not reported.

\section{Conclusion and Discussion}
We introduce view-centered P-FFN training with exact offline folding and the original ranking loss. CC-RankMixer improves AUC and LogLoss over RankMixer on three datasets, but regresses on Taobao-UB. The formal five-day online experiment reports +3.83\% duration per play and +7.41\% per user ($p=0.1169$ and $0.1575$), meeting the operational $p<0.20$ criterion but not $p<0.05$.

The SGD analysis describes local update geometry without establishing Adagrad's trajectory or better convergence. Matched controls, multiple seeds, gradient-coordinate verification, and online guardrail analysis remain necessary. Full-model folding equivalence requires the remaining scoring graph to stay unchanged.

\FloatBarrier
\begingroup\fontsize{10}{13.4}\selectfont

\endgroup

\end{document}